\documentclass[%
 aip,
 jmp,%
 amsmath,amssymb,
 reprint,%
]{revtex4-1}

\usepackage{graphicx}% Include figure files
\usepackage{dcolumn}% Align table columns on decimal point
\usepackage{bm}% bold math
\begin{document}

\preprint{AIP/123-QED}

\title[]{Water transport through functionalized nanotubes with tunable hydrophobicity}% Force line breaks with \\
%\thanks{Footnote to title of article.}

 \author{Ian Moskowitz}
\affiliation{Lehigh University, Department of Chemical Engineering, Bethlehem, PA 18015}
 \author{Mark A. Snyder}
 %\thanks{}
\affiliation{Lehigh University, Department of Chemical Engineering, Bethlehem, PA 18015}
 \author{Jeetain Mittal}
 \email[]{jeetain@lehigh.edu}
 %\homepage[]{www.lehigh.edu/~jem309}
 %\thanks{}
\affiliation{Lehigh University, Department of Chemical Engineering, Bethlehem, PA 18015}

\date{\today}% It is always \today, today,

\begin{abstract}
Molecular dynamics simulations are used to study the occupancy and flow of water 
through nanotubes comprised of hydrophobic and hydrophilic atoms, which are arranged 
on a honeycomb lattice to mimic functionalized carbon nanotubes (CNTs). We consider 
single-file motion of TIP3P water through narrow channels of (6,6) CNTs with varying 
fractions ($f$) of hydrophilic atoms. Various arrangements of hydrophilic atoms are 
used 
to create heterogeneous nanotubes with separate hydrophobic/hydrophilic domains 
along the tube as well as random mixtures of the two types of atoms. 
The water occupancy 
inside the nanotube channel is found to vary non-linearly as a function of $f$, and 
a small fraction of hydrophilic atoms ($f\approx 0.4$) are sufficient to induce 
spontaneous and continuous filling of the nanotube. Interestingly, the average 
number of water molecules inside the channel and water flux through the nanotube 
are less sensitive to the specific arrangement of hydrophilic atoms than to the fraction, $f$. Two different 
regimes are observed for the water flux dependence on $f$ -- an approximately linear 
increase in flux as a function of $f$ for $f<0.4$, and almost no change in flux for 
higher $f$ values, 
similar to the change in water occupancy. We are able to define an effective 
interaction strength between nanotube atoms and water's oxygen, based on a linear 
combination of interaction strengths between hydrophobic and hydrophilic nanotube 
atoms and water, that can quantitatively capture the observed behavior. 
\end{abstract}

\pacs{Valid PACS appear here}% PACS, the Physics and Astronomy
%Use showkeys class option if keyword
\maketitle

\section{\label{intro}Introduction} 
Fluid flow through nanoscopic pores of carbon 
nanotubes has received significant scientific attention owing to the potential 
applications that may result from the use of nanotubes in various 
technologies~\cite{Rasaiah:2007p3291,Whitby:2007p4607,Noy:2007p8657,Alexiadis:2008p8614}. 
Examples include nanotube-based membranes for water desalination and 
purification~\cite{kalra-2003,corry2008}, 
nanoscale transport for lab-on-a-chip devices~\cite{crevillen2009} and integrated 
circuits~\cite{cao2008}. 
At the fundamental level, fluid flow through nanotubes is endowed with various  
interesting phenomena -- breakdown of continuum 
hydrodynamics~\cite{Majumder:2005p4405,Thomas:2009p8819}, temperature induced 
hydrophobic-hydrophilic transitions~\cite{Wang-2008}, novel phase behavior~\cite{takaiwa2008}, etc. 
Moreover, nanotubes can be used as surrogates to study other complex flow systems such as biological channels like 
aquaporins~\cite{borgnia1999}. Therein, carbon nanotubes can be used as a simple 
prototype devoid of complexity present in the biological channels, but nonetheless 
capturing some important physical details~\cite{Zhu:2003}.  

Early molecular dynamics simulations by Hummer {\em et al.} showed that water can 
spontaneously fill the cavity of a (6,6) carbon nanotube and stay 
filled during the entire simulation time of 66 ns~\cite{Hummer:2001p8566}. 
They observed a pulse-like intermittent transmission of water molecules through the 
nanotube with similar rates as biological channels. The transition between empty and 
filled nanotubes, realized by tuning interactions between nanotube atoms and 
water~\cite{Melillo-2011}, 
can be described as a two-state process~\cite{waghe:10789}.    
Several theoretical models have been developed that can describe water 
permeation through the nanotube channel at equilibrium~\cite{Berezhkovskii:2002p153} 
as well as in the presence of a driving force such as pressure or chemical 
potential difference~\cite{Zhu:2004p8945}. 

%Recent simulations using different diameter $D$ carbon nanotubes (CNTs) found a 
%flow transition from subcontinumm to continuum transport for $D\approx 1.39$ nm 
%in contrast to experimental results showing a breakdown of continuum transport 
%for $D=7$ nm. As an ultimate goal is to be able to pump water through these 
%nanochannels and several interesting ideas have been put forward such as 
%use of charges positioned along the nanotube and use of rotation-translation 
%coupling between neighboring water molecules for water pumping. 

While simulations on pristine CNTs have provided invaluable fundamental insight into water 
flow therein, functionalization of carbon nanotubes~\cite{banerjee2003} is often either a prerequisite 
for their 
processing, e.g., the use of surfactant for solution dispersion~\cite{moore2003}, or 
desirable to achieve certain functionality such as sensitivity and selectivity in 
molecular detection~\cite{qi2003}. 
%an important question. 
Moreover, nanotubes with similar structural features as CNTs, but different atomic 
composition such as Boron-Nitride~\cite{chopra1995}, carbon doped 
Boron-Nitride~\cite{wei2010}, 
and various inorganic materials~\cite{remvskar2004}, can now be synthesized routinely 
in the laboratory. 
In the context of CNT-surrogate study of flow through biological channels, nanotube-water interactions can vary depending 
on the hydrophobicity of the amino acids involved. 
In all of these cases, as the interactions 
between nanotube atoms and water will change depending on the nanotube 
functionalization and composition, the water filling and permeation behavior will also 
be different.  
While the understanding of water flow through functionalized nanotubes so far is quite 
limited, simple models that incorporate 
basic details of nanotube functionalization may enable important progress 
in understanding fluid flow through nanotubes of a broader range of composition and function.

In this paper, we use molecular dynamics simulations to study water flow 
through model tubular channels in which atoms are arranged on a 
honeycomb lattice, and two different atom types are used to populate the lattice 
sites, resulting in a binary heterogeneous nanotube. 
The two types of nanotube atoms differ only in their Lennard-Jones 
parameters with respect to water, and correspond to sp$^2$-hybridized carbon $C$ (labeled 
hydrophilic due to continued filling of the nanotube with water) and reduced 
carbon-water interaction strength $C^{\prime}$ (termed hydrophobic as the nanotube is mostly 
empty but undergoes transition between empty and filled states) as used by 
Hummer {\em et al}.~\cite{Hummer:2001p8566}. 
We note that our classification of nonpolar sp$^2$-hybridized atoms as hydrophilic is 
different from previous work where the nanotube made up of such atoms was actually characterized 
as hydrophobic~\cite{Hummer:2001p8566}. We have based our classification on the water droplet 
contact angle on a flat unrolled nanotube surface; for $C$ atoms the contact angle is less than 
90$^o$ (hydrophilic) and for $C^{\prime}$ atoms the contact angle if slightly above 90$^o$ (hydrophobic)~\cite{Melillo-2011}.
We first vary the fraction, $f$, of hydrophilic nanotube atoms, separated axially from the hydrophobic 
nanotube atoms, to study its impact on water filling as well as water flow through the nanotube. In addition, for a 
given $f$, we also study the influence of 
nanoscale patterning on water flow by arranging hydrophilic nanotube atoms in various patterns. Finally, we propose a simple mean-field model, in which the heterogeneous 
nanotube is represented as a homogeneous nanotube with uniform interactions between the nanotube atoms and water. 
We show that the predictions of this simple model are in very good agreement with the detailed simulation data.
%We ask the following 
%questions: (i) how do the nanotube composition defined by fraction of sites 
%occupied by $C$ atoms, $f$, affects hydration and equilibrium flux?, 
%(ii) how sensitive are the results to a precise arrangement of $C$ vs $C^{\prime}$ 
%atoms on a tubular lattice?, and (iii) if these results offer an insight 
%useful for the design of nanoscale transport devices.    

\section{Model and Simulation Details}
To study the flow of water through functionalized nanotubes, a single (6,6) 
nanotube, with diameter of 0.81 nm and length 1.34 nm, is solvated in a periodic TIP3P water 
box with 1025 molecules. The nanotube is free to translate and rotate througout the simulation 
time. The simulations are performed in an NPT ensemble using  
molecular simulation package GROMACS (version 4.0.7) for 100 ns with a timestep of 2 fs. 
The temperature is maintained at 300 K using the Nose-Hoover thermostat, and the pressure is 
maintained at 1 bar using the Parrinello-Rahman barostat with a time constant of 5 ps. 
We use the particle mesh Ewald (PME) method~\cite{essmann1995} for long-range electrostatic 
interactions with a short-range cutoff value of 0.9 nm. For short-range Lennard-Jones (LJ) interactions, 
we use a cutoff value of 1.4 nm. The well-depths of the Lennard-Jones potential
between the nanotube atoms and water's oxygen atom are 0.4784 kJ/mol (hydrophillic) and 
0.2703 kJ/mol (hydrophobic), with the corresponding LJ diameters being 0.3275 nm (hydrophillic) 
and 0.3414 nm (hydrophobic). The angle and dihedral potentials are also applied to the nanotube atoms 
as done previously~\cite{Hummer:2001p8566}. 

\section{Results and Discussion}

\begin{figure*}[ht]
\begin{tabular}{ccc}
%\scalebox{1.00}{\includegraphics*[width = 6.5 cm]{Figure1a.pdf}} &
\scalebox{1.00}{\includegraphics*[height = 5.0 cm]{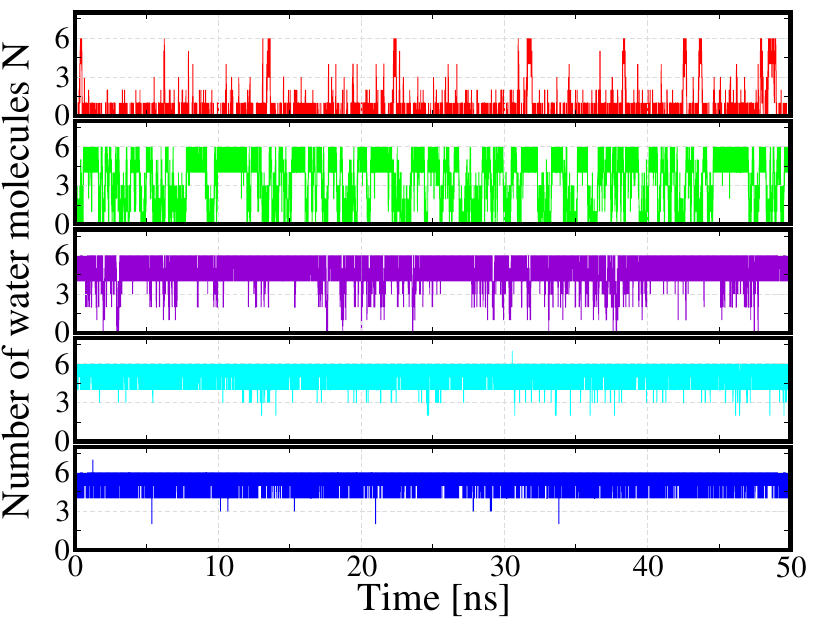}} &
\scalebox{1.00}{\includegraphics*[height = 1.90 in]{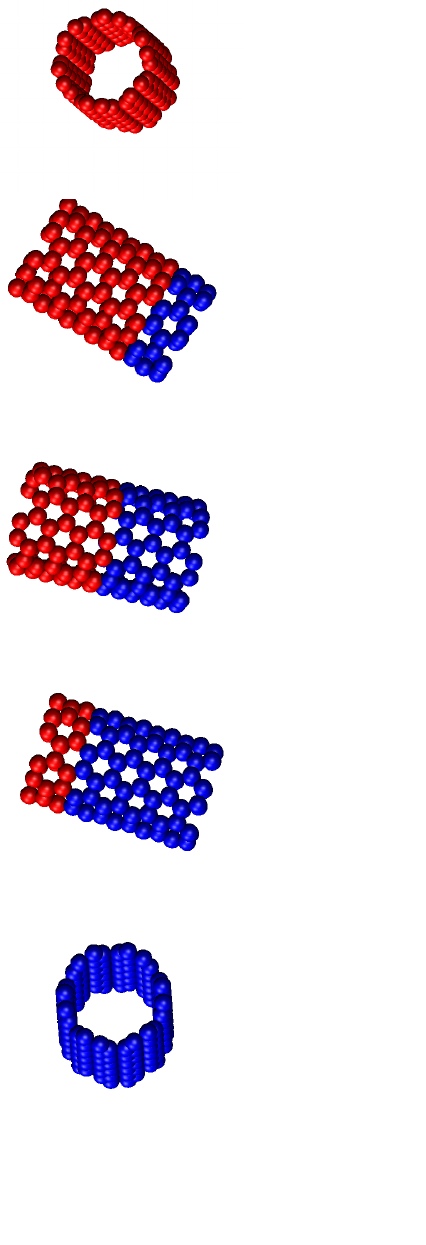}} &
\scalebox{1.00}{\includegraphics*[height = 5.0 cm]{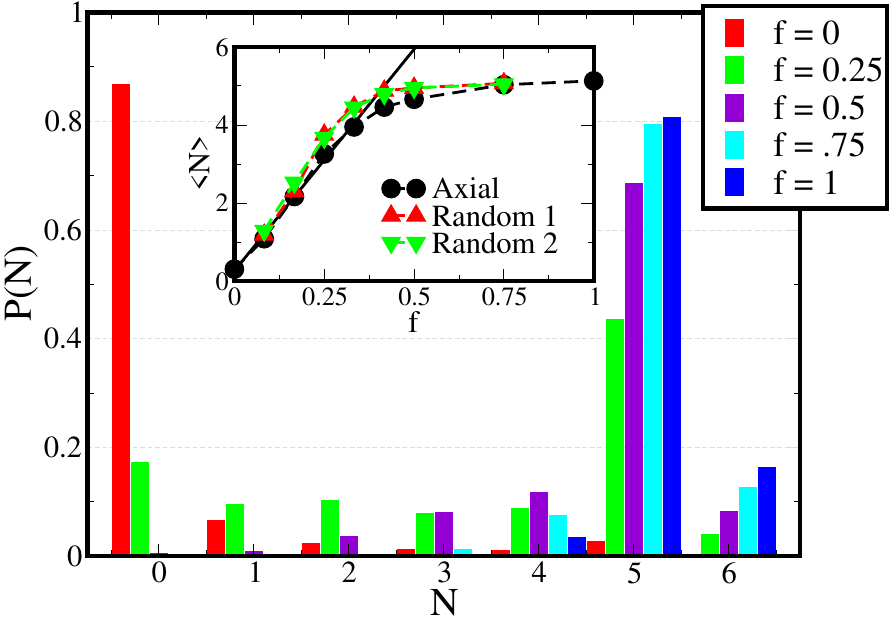}} \\
\end{tabular}
\caption{Water occupancy. (Left panel) Number of water
molecules $N$ for varying fraction of hydrophilic nanotube atoms $f$
are shown as a function of time. (Middle panel) Nanotube snapshots are shown, with 
hydrophobic and hydrophilic atoms labeled in red and blue, respectively. (Right panel) Probability of observing $N$ water molecules inside the nanotube $P(N)$ is shown for various values of $f$. The inset shows average number of water molecules as a function 
of $f$ for three possible arrangements of hydrophilic atoms. 
}
\label{fig1}
\end{figure*}

To understand the dependence of water flow through cylindrical channels with both hydrophobic and 
hydrophilic building blocks, we use carbon nanotube like short cylindrical channels made up of 
hydrophobic and hydrophilic atoms. The fraction of hydrophilic atoms ($f$) is varied to tune 
the relative proportions of these atoms in order to change the water occupancy and flow characteristics of 
the resulting nanotubes. Previously, Hummer {\em et al.} have shown that fully hydrophilic 
nanotubes ($f=1$) of the same size stay filled with an average occupancy of about 5 water molecules, and conduct 
about 17 water molecules per nanosecond~\cite{Hummer:2001p8566}. 
On the other hand, fully hydrophobic nanotubes ($f=0$) 
can transition between filled and empty states with an apparent two-state kinetics~\cite{waghe:10789}. 
Melillo {\em et al.} performed a systematic variation of interaction strength between nanotube 
atoms and water's oxygen to show that water occupancy and flow is sensitive to changes in the 
interaction strength within only a narrow range~\cite{Melillo-2011}. For low attraction strengths, 
the nanotubes  
stay empty with no flow, whereas for high attraction strengths, the nanotubes are filled and 
conduct water with similar rates. For intermediate attraction strengths, a sharp increase in 
water occupancy and flow was observed, which was related to the change in wetting 
characteristics based on simulations carried out on the unrolled nanotube surface. 

Recent work from Debenedetti and co-workers 
has highlighted the importance of surface heterogeneity and patterning on nanoscale hydrophobicity 
as it relates to the behavior of confined water~\cite{giovambattista2007hydration,Giovambattista:2009p8697}. 
Some of the complexities associated with water flow through heterogeneous channels have been addressed using charged 
nanotube atoms~\cite{Zhu:2003}, asymmetrically positioned charges~\cite{Zuo:2009p8951}, and Y-shaped 
nanochannels~\cite{Tu:2009p8910}. 
For nanotube surfaces comprised of hydrophilic and hydrophobic atoms, the sensitivity of water flow to 
the corresponding surface heterogeneity, the fraction of hydrophilic atoms, and the actual patterning of these atoms into hydrophobic and hydrophilic domains, 
is much less clear. 

Figure~\ref{fig1} (left panel) shows the number of water molecules occupying the nanotube 
as a function of time for varying fractions of hydrophilic atoms, $f=0$, 0.25, 0.5, 0.75, and 1 
(from top to bottom). In this case, hydrophilic and hydrophobic atoms are distributed along the 
nanotube axis termed `axial arrangement' to distinguish it from other arrangements considered in 
this work. 
As expected on the basis of earlier findings, the fully hydrophobic nanotube ($f=0$) undergoes transitions 
between empty and filled states, whereas the fully hydrophilic nanotube ($f=1$) stays filled during 
the entire simulation time. Interestingly, we find that just a small fraction of hydrophilic atoms ($f=0.25$), localized at the 
nanotube tip, is sufficient to significantly increase the number of water molecules occupying the 
nanotube. For $f>0.4$, the nanotube stays completely filled. 

To further quantify water 
occupancy as a function of $f$, we calculate the distribution $P(N)$, which is the probability of 
finding $N$ water molecules inside the nanotube. The $P(N)$ for several representative values of $f$ are shown in 
Fig.~\ref{fig1} (right panel). The probability of observing an empty nanotube, $P(0)$, decreases 
dramatically with increasing number of hydrophilic nanotube atoms (i.e., increasing $f$ to $0.5$). 
This precipitous decrease in $P(0)$ is complemented by a similarly dramatic rise in the probability of observing $N=5$ or 6 water molecules, 
whereas relatively little change in 
$P(N)$ is observed with subsequent increase in $f$ from $f=0.75$ to $f=1$. 
%In this work, we have considered nanotubes with 
%small diameters that can only accomodate water molecules in a single-file, thereby %limiting  
%significant enhancement in water filling with changes in system parameters such as %favorable 
%nanotube-water interactions~\cite{Melillo-2011}.  
As shown in the inset of Fig.~\ref{fig1} 
(right panel), the average number of water molecules inside the nanotube $<N>$ changes approximately 
linearly as a function of $f$ for $f<0.4$ and then plateaus at a value found for the fully hydrophilic case.
The so-called 'axial arrangement' is the same as depicted in 
Fig.~\ref{fig1}, which is obtained by separating continuous domains of hydrophilic and hydrophobic atoms along the nanotube 
axis. The so-called `random arrangement' is obtained by placing the hydrophilic atoms randomly on the nanotube 
lattice (lattice sites numbered from 1 to 144), with two different cases considered here by 
changing the seed of the random number generator. 

\begin{figure}
\centering
\scalebox{1.00}{\includegraphics*[width = 8 cm]{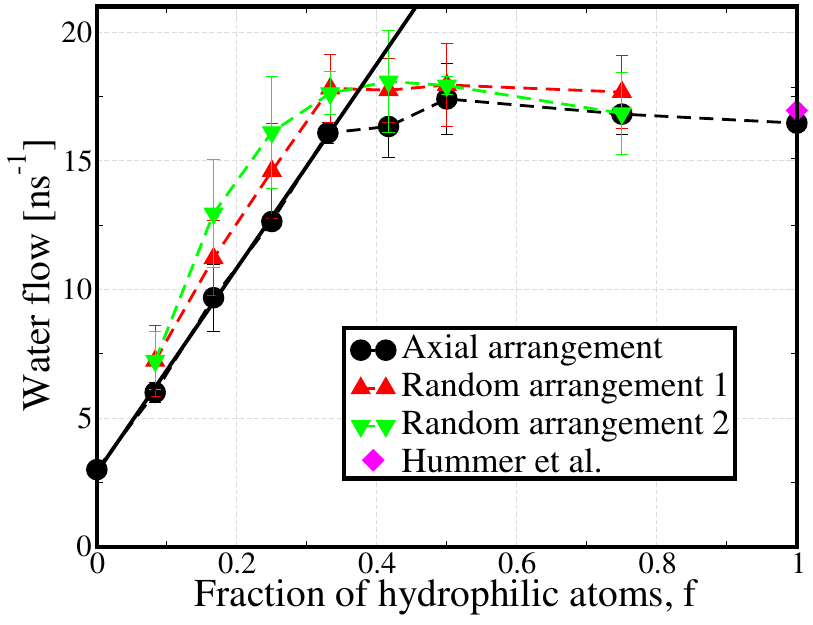}}
\caption{Water flow rate. Number of water molecules permeating through the nanotube, which enter either side and leave from the opposite side, per unit time, are shown as a
function of fraction of hydrophilic atoms $f$. The symbols connected by dashed lines are the simulation data and the solid black line is the linear fit for $f<0.4$ for the `axial 
arrangement' data.}
\label{fig2}
\end{figure}

In addition to water occupancy, water transport and its potential sensitivity to tube properties, 
are important to quantify. To probe the effects of changing the fraction of hydrophilic atoms on water 
flux through the nanotube, we calculate the number of water molecules that permeate the 
nanotube by entering either end, but leaving from the opposite end over the total simulation 
time. Water flow is defined as the number of water molecules transmitted per nanosecond 
through the nanotube. Figure~\ref{fig2} shows water flow as a function of $f$ for three possible 
arrangements of hydrophilic nanotube atoms. 
For reference, we show the value of water flow 
obtained by Hummer {\em et al.}~\cite{Hummer:2001p8566} for the fully hydrophilic nanotube ($f=1$). 

First, focusing on the data for the axial arrangement of hydrophilic and hydrophobic atoms in Fig.~\ref{fig2}, we observe an approximately 
linear dependence of water flow on $f$ for $f<0.4$. Outside this regime, the water flow is 
relatively insensitive to the increase in $f$, and a slight decrease in water flux is observed with 
increasing $f$. The linear dependence of water flow  as a function of $f$ for $f<0.4$ likely appears due to the concomitant  
linear relationship between water flow and occupancy, shown for the same condition in Fig.~\ref{fig1} (inset). 
A similar change in water occupancy and 
flow was reported previously for a spatially uniform change in nanotube-water attraction strength over low to 
moderate interaction strengths~\cite{Melillo-2011}. There, an increase in the nanotube-water attraction 
strength was shown to monotonically increase water's residence time inside the nanotube. 
For higher attraction strengths, this increase in the residence time led to a decrease in water flow, thereby deeming 
both 
the average water occupancy and the residence time important for capturing the change in 
water flow. In the case of heterogeneous nanotubes, an increase in the fraction of hydrophilic atoms 
above $f=0.4$ leads to a slight increase in water occupancy (Fig.~\ref{fig1}), but insensitive or subtly decreasing water 
flow (Fig.~\ref{fig2}). Therefore, the water flow characteristics appear to be influenced similarly by a heterogeneous 
distribution of hydrophilic and hydrophobic atoms as compared to a spatially uniform change in 
nanotube-water interaction strength.     

To see if the specific axial arrangement of hydrophilic atoms is responsible for the observed 
behavior, Fig.~\ref{fig2} also shows water flow as a function of $f$ for random arrangements of 
hydrophilic atoms. The qualitative behavior is quite similar in all cases, with approximately 
linear dependence for low $f$ values and a plataeu (or slight decrease) for higher $f$ values. 

Though statistically less significant, a slight increase in water flow for a given fraction 
of hydrophilic atoms is observed in the case of the random arrangement as opposed to the axial arrangement. 
The water occupancy data (Fig. 1, right panel inset) also shows a similar trend, with the random arrangement leading to 
slightly higher water occupancy. One can explain this based on previous work by Giovambattista {\em et al.} 
where it was shown that hydrophilic borders surrounding nanoscale hydrophobic patches 
can significantly alter water hydration by reducing the water repulsion~\cite{giovambattista2007hydration}. 
In the `random 
arrangement' case, the hydrophilic atoms are mixed with hydrophobic atoms, thereby enhancing water 
density near such surfaces, which can lead to slightly enhanced water flow. %Based on our previous 
%work, we expect this effect to become stronger with increasing nanotube diameter and length.  
 
\begin{figure}
\scalebox{1.00}{\includegraphics*[width = 8 cm]{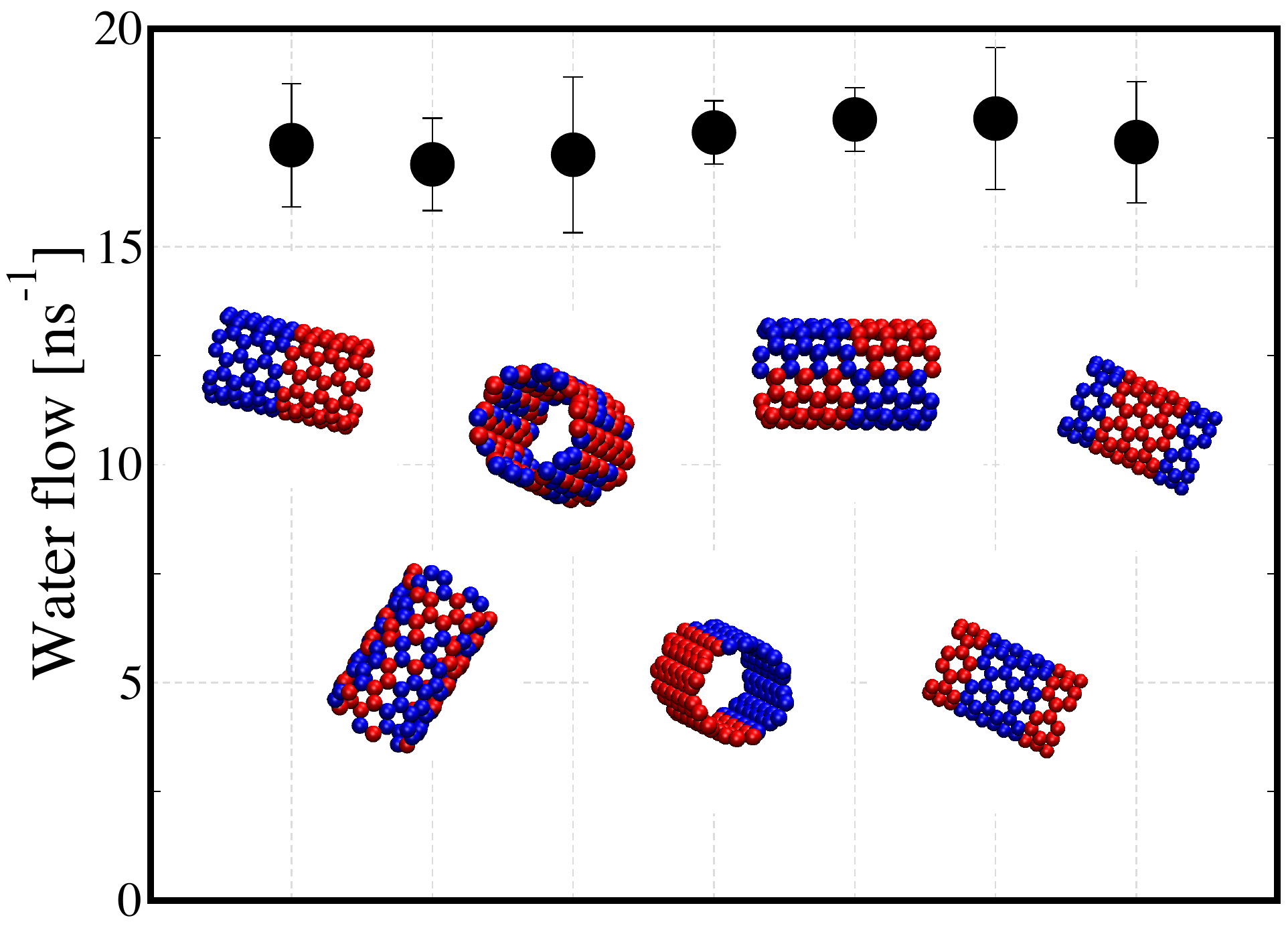}}
\caption{Number of water molecules permeating through the nanotube, which enter either side and leave from the opposite side, per unit time are shown for various arrangements of hydrophilic (blue) and hydrophobic atoms (red) for $f=0.5$ as depicted in the nanotube snapshots.}
\label{fig3}
\end{figure}

Between the cases of axial and random arrangement of hydrophilic and hydrophobic atoms exists a range of cases wherein
discrete domains could conceivably pattern the nanotube surface. 
To further probe how such specific arrangement of hydrophilic atoms (e.g., on the ends of the nanotube) 
may modulate water flow, we simulate several possible arrangements as shown in 
Fig.~\ref{fig3}, all for $f=0.5$. Within statistical uncertainty of the measured water flow rate, the 
different heterogeneities studied produce similar results. As the water flow through these short nanotubes is 
expected to be dominated by the barriers at the entry and exit of the nanotube~\cite{kalra-2003}, the 
observed insensitivity of water 
flow to the actual arrangement of hydrophilic atoms is a bit surprising.  
While this effect may be tuned by increasing nanotube length and for low values of $f$, this requires future work employing either a low-cost but accurate single-site water model~\cite{molinero2008water} 
or dipole lattice model of water~\cite{kofinger2008}.     

\begin{figure}
\centering
\scalebox{1.00}{\includegraphics*[width = 6 cm]{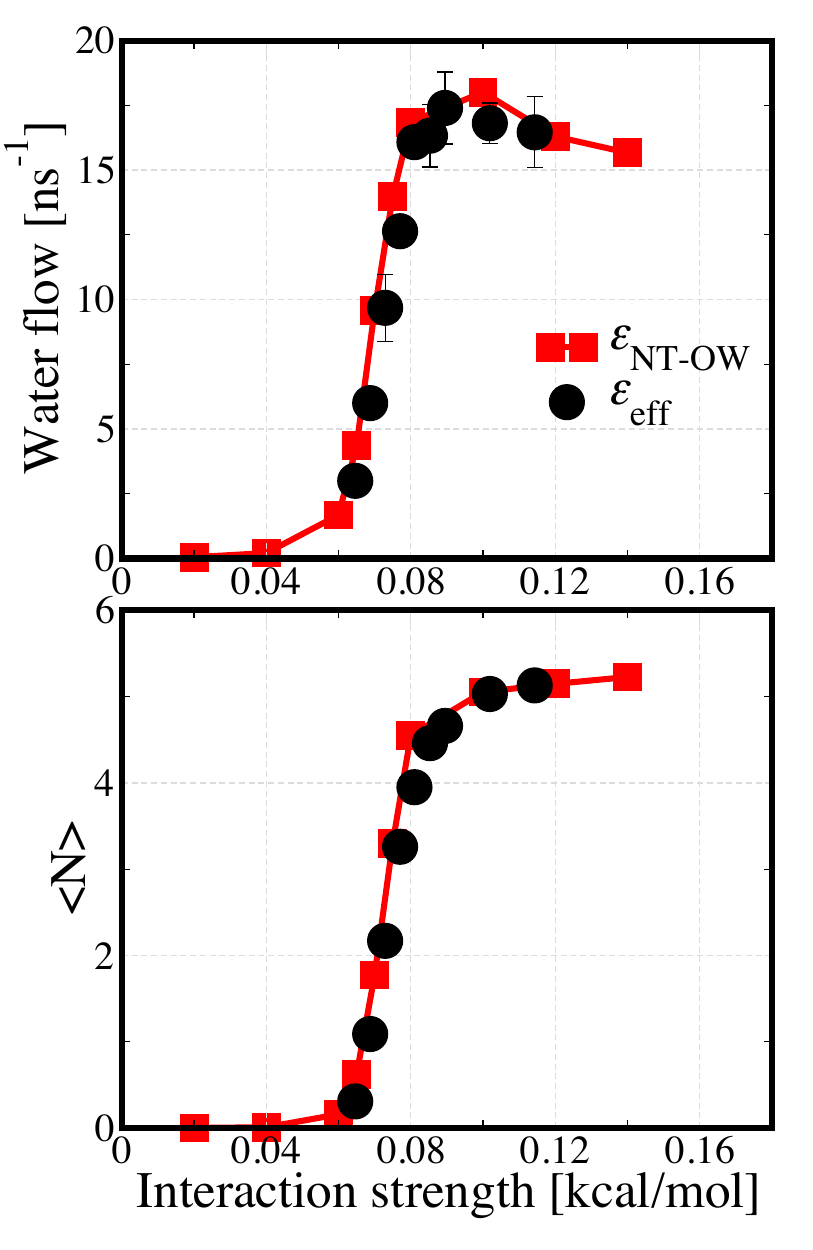}}
\caption{Flow through homogeneous versus heterogeneous nanotubes. (Top panel) Number of water 
molecules permeating through the nanotube, which enter either side and leave from the opposite side, per unit time are shown as a
function of interaction strength between nanotube atoms and water's oxygen. For 
nanotubes comprised of hydrophobic and hydrophilic atoms, we use eq. 1 to estimate 
an effective interaction strength $\epsilon_{\mathrm{eff}}$, whereas separate simulations 
are conducted for nanotubes comprised of a single type of atoms interacting with water's 
oxygen with $\epsilon_{\mathrm{NT-OW}}$. (Bottom panel) Average number of water molecules as a 
function of interaction strength between nanotube atoms and water's oxygen. 
}
\label{fig4}
\end{figure}

The qualitative similarity we observe here in water flow through heterogenous nanotubes made up 
of two types of atoms with previous work on homogeneous scaling of nanotube-water interaction 
strength, begs the question about whether these two cases may be related in a more quantitative way. 
Due to the observed insensitivity of the water occupancy and flow on the actual arrangement of 
the hydrophilic atoms, we {\em hypothesize} that one can define an effective interaction strength 
$\epsilon_{\mathrm {eff}}$ between nanotube atoms and water's oxygen as, 

\begin{equation}
\epsilon_{\mathrm {eff}} = \epsilon_{\mathrm {philic}}  f + \epsilon_{\mathrm {phobic}}  (1 - f), 
\end{equation}

where $\epsilon_{\mathrm {philic}}$ and $\epsilon_{\mathrm {phobic}}$ correspond to the LJ 
interaction parameters between nanotube atoms and water's oxygen for hydrophilic and hydrophobic 
atoms, respectively. Such a linear interpolation is also justified based on the small differences in the 
interaction potential curves (LJ potential) between the two cases~\cite{Hummer:2001p8566}. 
To test if such a simple scaling relationship works over a broad range, we 
calculate water flow through a nanotube and average water occupancy for various nanotube-water interaction strengths 
$\epsilon_{\mathrm {NT-OW}}$. The comparison of this data with water flow data from 
Fig.~\ref{fig2} and water occupancy data from Fig.~\ref{fig1} (right panel inset) is shown in Fig.~\ref{fig4}. 
Overall, the change in water flow and $<N>$ as a function of 
uniform change in nanotube-water interaction strength, as predicted by detailed simulations, is in remarkably good agreement with the prediction based on 
the mean-field type description in 
Eq. 1 for heterogeneous nanotubes. This quantitative agreement in Fig.~\ref{fig4} suggests that the 
fluid flow through functionalized nanotubes may be a simple function of a parameter describing 
the affinity between the fluid molecules and the nanotube surface.  
The apparent insensitivity to the distribution of the two atom types may be a manifestation of the 
single-file nature of water flow through a (6,6) nanotube. This highlights the question for future 
study regarding whether deviations from a mean-field description become significant for larger 
diameter nanotubes or not.
In addition, it will be useful for future studies to elucidate any sensitivity of the observed behavior 
to nanotube length.

%one would like to 
%understand if the nanotube length plays a significant role in the observed behavior or not. 
%We note that it is not entirely surprising in this case given that the water flow happens in a single-file 
%manner through a (6,6) nanotube. It will be interesting to see, in future, if the deviations from a 
%mean-field description are significant for larger diameter nanotubes or not. 

\section{Conclusions}

We have studied the water filling and transport through a simple model of functionalized carbon nanotubes. 
The fraction of hydrophilic atoms, $f$, are varied to change the nanotube from fully hydrophobic ($f=0$) to 
fully hydrophilic ($f=1$). Two different regimes are identified for water occupancy and flow as a function of  
$f$. For low $f$ values ($f<0.4$), there is an approximately linear dependence on $f$ of the average number of water 
molecules inside the nanotube as well as the number of water molecules transmitting through the nanotube. For higher $f$ values, 
a slight increase in water occupancy is observed, whereas a slight decrease in water 
flow is predicted. For most of the systems considered in this work, the observed water flow properties are relatively 
insensitive to the precise arrangement of hydrophilic atoms. A simple mean-field 
type expression for effective interaction strength between nanotube atoms and water can 
be used to quantitatively capture flow characteristics of water through heterogeneous nanotubes. These findings suggest that control over
the character (i.e., interactions with water) of the nanotube functionality may be more important than the specific
spatial arrangement of that function for controlling water flow therein.  This simple design principle 
should enable prediction of flow characteristics in synthetic and functionalized nanotubes
beyond pristine CNT and help guide rational design of target functionality.

\section{acknowledgments}
This work was supported by the National Science Foundation grant CBET-1120399.  Use of the high-performance computing capabilities of the Extreme Science and Engineering Discovery Environment (XSEDE), which is supported by the National Science Foundation grant no. TG-MCB-120014, is also gratefully acknowledged.
%\end{acknowledgments}

%\bibliography{all}
%
% ****** End of file aipsamp.tex ******

\end{document}